\documentclass[preprint,sort&compress]{elsarticle}
\usepackage{placeins}

\usepackage{amssymb,amsthm,amsmath,caption,commath,bbm,psfrag,verbatim}
\usepackage{setspace,color,url,multirow,array,graphicx,epstopdf,hyperref,float}
\usepackage{tikz}
\usepackage[FIGTOPCAP,nooneline]{subfigure}
\usepackage{capt-of}

\DeclareMathOperator{\Real}{Re}

\newcommand{\calR}{\mathcal{R}}
\newcommand{\R}{\mathbb{R}}
\newcommand{\calJ}{\mathcal{J}}
\newcommand{\calC}{\mathcal{C}}

\makeatletter

\newcommand{\Rom}[1]{\expandafter\@slowromancap\romannumeral #1@}
\makeatother

\title{An SIRS-model considering waning efficiency and periodic re-vaccination}

\author[ESPOL,CFD]{Joseph P\'aez Ch\'avez}
\ead{jpaez@espol.edu.ec}

\author[ODU]{Ayt\"ul G\"ok\c{c}e}
\ead{aytulgokce@odu.edu.tr}

\author[KOB]{Thomas G\"otz}
\ead{goetz@uni-koblenz.de}

\author[MAI,IQCB]{Burcu G\"urb\"uz}
\ead{burcu.gurbuz@uni-mainz.de}

\address[ESPOL]{Center for Applied Dynamical Systems and Computational Methods (CADSCOM), Faculty of Natural Sciences and Mathematics, Escuela Superior Polit\'ecnica del
Litoral, P.O. Box 09-01-5863, Guayaquil, Ecuador}

\address[ODU]{Ordu University, Faculty of Science and Letters, Department of Mathematics, 52200 Ordu, Turkey}

\address[KOB]{Mathematical Institute, University of Koblenz, D-56070 Koblenz, Germany}

\address[MAI]{Institut f\"ur Mathematik, Johannes Gutenberg-Universit\"at Mainz, Staudingerweg 9, 55099 Mainz, Germany}

\address[CFD]{Center for Dynamics, Department of Mathematics, TU Dresden, D-01062 Dresden, Germany}

\address[IQCB]{Institute for Quantitative and Computational Biosciences (IQCB), Johannes Gutenberg University Mainz, 55128 Mainz, Germany}

\begin{document}

\begin{abstract}

In this paper, we extend the classical SIRS (Susceptible-Infectious-Recovered-Susceptible) model from mathematical epidemiology by incorporating a vaccinated compartment, V, accounting for an imperfect vaccine with waning efficacy over time. The SIRSV-model divides the population into four compartments and introduces periodic re-vaccination for waning immunity. The efficiency of the vaccine is assumed to decay with the time passed since the vaccination. Periodic re-vaccinations are applied to the population. We develop a partial differential equation (PDE) model for the continuous vaccination time and a coupled ordinary differential equation (ODE) system when discretizing the vaccination period. We analyze the equilibria of the ODE model and investigate the linear stability of the disease-free equilibrium (DFE). Furthermore, we explore an optimization framework where vaccination rate, re-vaccination time, and non-pharmaceutical interventions (NPIs) are control variables to minimize infection levels. The optimization objective is defined using different norm-based measures of infected individuals.
A numerical analysis of the model's dynamic behavior under varying control parameters is conducted using path-following methods. The analysis focuses on the impacts of vaccination strategies and contact limitation measures.
Bifurcation analysis reveals complex behaviors, including bistability, fold bifurcations, forward and backward bifurcations, highlighting the need for combined vaccination and contact control strategies to manage disease spread effectively.
\end{abstract}


\maketitle
\section{Introduction}
\label{Sec:Intro}
Mathematical models have been essential to understanding the spread and control of infectious diseases, with the classical SIR (Susceptible-Infectious-Recovered) model being one of the most widely used frameworks in epidemiology. This model has been fundamental in analyzing the transmission dynamics of many infectious diseases by partitioning the population into susceptible, infected, and recovered individuals~\cite{KerMcK27}. However, several complexities of the real world, such as waning immunity and vaccine dynamics, require extensions to the standard SIR-model to capture the full range of disease transmission and control efforts. Specifically, accounting for waning vaccine efficiency and periodic re-vaccination is critical to designing effective immunization strategies, especially for diseases where immunity is not permanent \cite{anderson91}, \cite{hethcote00}. For diseases such as pertussis, influenza, and COVID-19, waning immunity following natural infection or vaccination has been previously studied \cite{plotkin10}. Such waning immunity can lead to a growing pool of susceptible individuals over time, increasing the likelihood of subsequent outbreaks. To address this, many public health strategies include periodic re-vaccination, especially for vaccines that do not provide lifelong immunity. Incorporating these factors into an SIR-based framework provides a more realistic representation of disease dynamics, especially in populations undergoing periodic vaccination campaigns \cite{Doenges24}, \cite{gokcce2024dynamics}.

On the other hand, SIR-based models are used to develop new mathematical models for complex systems. In recent years, the SIR-model has been extended in various ways to address the complexity of infectious diseases with vaccination and waning immunity. Ehrhardt et al. (2019) present a comprehensive SIR-based model that accounts for both vaccination and waning immunity, highlighting the importance of incorporating these factors for a realistic representation of disease dynamics \cite{Ehrhardt19}. Sun and Yang further explore global outcomes in an SIRS-model, examining the effects of vaccination and isolation on disease control and demonstrating that isolation can significantly reduce infection rates \cite{Sun2010}. González and Villena (2010) analyze the spatial dynamics of vaccination using a spatial SIRS-V model, emphasizing how spatial factors influence the spread of infectious diseases and vaccination strategies \cite{Gonzalez20}. In addition, Elbasha et al. (2011) explore the dynamics of an SIRS-model that incorporates natural and vaccine-induced immunity, highlighting the implications of waning immunity for disease re-emergence and vaccination efforts \cite{Elbasha11}. Barbarossa and Röst (2015) examined immune status-structured populations, providing a detailed immuno-epidemiological perspective on how waning immunity and immune boosting can shape epidemic dynamics, highlighting the importance of immune memory and boosting events in disease management \cite{Barbarossa15}. Together, these studies highlight the necessity of incorporating vaccination, waning immunity, and spatial considerations into epidemiological models to improve disease management strategies \cite{bjornstad20}.

In this paper, we study a modified model as an extension of the classical SIRS-model that incorporates both waning vaccine efficacy and periodic re-vaccination, which are critical for managing diseases with incomplete or time-limited immunity. We introduce a the vaccination age $x$ as the time elapsed since the (last) vaccination of an individual. The vaccinated compartment itself then depends both on time and vaccination age. The declining efficacy of the vaccine is modeled by a decreasing function that reflects the gradual reduction in immune protection conferred by vaccination \cite{wearing09}. Individuals in the vaccinated compartment are periodically re-vaccinated, which helps to maintain immunity in the population.
The present extension enables to analyze both the direct effects of vaccination as well as the relationship between vaccination schedules, waning immunity, and the subsequent impact on the transmission dynamics of disease.

The remainder of this paper is organized as follows. In Section 2, the mathematical formulation of the SIRSV-model with continuous vaccination age is presented. We briefly discuss the equilibria of this PDE model showing a parameter window where two endemic equilibria may co-exist. In Section 3 we discretize the vaccination age to reduce the model to a coupled, but high-dimensional ODE system. Again, we can discuss its equilibria that agree with the ones of the PDE model.
The linear stability of the disease-free equilibrium is analyzed, and the basic reproduction number of the model is computed.
In Section 4, we introduce the numerical continuation techniques used to explore the parameter space and investigate the impact of NPIs and we present numerical simulations of various disease control strategies, highlighting the effectiveness of combined vaccination and contact restriction policies. Finally,
Section 5 is devoted to a discussion of the implications of the findings and the future directions of research.

\section{SIRSV--PDE-model with continuous vaccination age}
\label{Sec:Model}
We consider a classical SIRS-model from mathematical epidemiology and extend it introducing a vaccinated compartment. The entire population $N$ is subdivided into the susceptible compartment $S$, the infected compartment $I$, the recovered compartment $R$ and the vaccinated compartment $V$. Susceptible individuals get infected with rate $\beta$ when in contact with an infected. Infected individuals recover with rate $\gamma$ and recovered individuals loose their immunity with rate $\alpha$ and become susceptible again. For the vaccinated individuals, we introduce the time $x$ elapsed since vaccination, called the \emph{vaccination age}. The vaccine is assumed to be \emph{imperfect} and its waning efficiency is modeled by a decreasing function $\omega(x)$ depending on the vaccination age $x$. The probability, that a vaccinated individual gets infected is given by $1-\omega$. To partially compensate for the waning efficiency, we assume periodic update vaccinations, i.e.~individuals that have been vaccinated $P$ time units ago, receive a new vaccination.

Modeling the vaccination age $x$ as a continuous variable, we arrive at a PDE model.
\begin{subequations}
\begin{align}
    \frac{dS}{dt} &= - \beta S \frac{I}{N} - \nu S + \alpha R, \label{E:PDE:S}\\
    \frac{dI}{dt} &= \beta \frac{I}{N} \left( S + \int_0^P (1-\omega(x)) V(t,x)\, dx \right) - \gamma I, \label{E:I1}\\
    \frac{dR}{dt} &= \gamma I -\alpha R, \label{E:PDE:R}
\intertext{and a transport through time and vaccination age}
    \frac{\partial V}{\partial t} + \frac{\partial V}{\partial x} &= -\beta \frac{I}{N} (1-\omega(x)) V(t,x).
    \label{E:TranspV}
\intertext{The re-vaccination at time $x=P$ leads to the
conditions}
    V(t,0) &= \nu S + V(t,P^-), \label{E:VP} \\
    V(t,P^+) &= 0.   \label{E:ReVacc2}
\end{align}
\end{subequations}
Here $V(t,P^-):=\lim\limits_{x\nearrow P^-} V(t,x)$ denotes the left-sided limit of $V(t,\cdot)$ at $x=P$ and $V(t,P^+)$ analogously denotes the right-sided limit.

If NPIs are applied to complement the vaccination campaign, we assume that these NPIs affect the transmission rate $\beta$, i.e.
\begin{align}
\label{E:NPI}
    \beta(t) := \beta_0 \left( 1- \rho(t) \right),
\end{align}
where $\rho\in [0,1]$ denotes the severity of the restrictions,
$\rho=0$ means no restrictions and $\rho=1$ means total lockdown.

To compute (temporal) equilibria of the PDE model, we assume $\omega(x)=\omega$ to be independent of the vaccination age $x$.
The Eqns.~\eqref{E:PDE:R} and~\eqref{E:PDE:S} yield $R^\ast=\frac{\gamma}{\alpha} I^\ast$ and $S^\ast = \frac{\gamma N}{\beta I^\ast+\nu N} I^\ast$. The transport Eqn.~\eqref{E:TranspV} reads in the temporal equilibrium as $\partial_x V^\ast = - \frac{\beta (1-\omega)}{N} I^\ast V^\ast$ and hence
\begin{equation*}
    V^\ast(x) = V_0^\ast \exp \left[ -\frac{(1-\omega)\beta}{N} I^\ast x\right]\;.
\end{equation*}
In the infection equation~\eqref{E:I1}, the integral runs effectively only over the interval $0\le x\le P$, since at time $P$, individuals get re-vaccinated and hence there are no indiviuduals with vaccination age larger than $P$. We obtain in the equilibrium either the trivial solution $I^\ast=0$ or
\begin{gather*}
    \frac{\beta}{N} \left( \frac{\gamma N I^\ast}{\beta I^\ast+\nu N} + V_0^\ast(1-\omega) \int_0^P \exp \left[ -\frac{(1-\omega)\beta}{N} I^\ast x\right]\, dx\right) - \gamma = 0,
\intertext{and after equating the integral and resorting terms, we arrive at}
    \frac{\beta\gamma I^\ast}{\beta I^\ast+\nu N}
    +\frac{V_0}{I^\ast} \left( 1- \exp\left[-\frac{\beta(1-\omega)P}{N}I^\ast\right]\right)- \gamma = 0,
\end{gather*}
and finally
\begin{equation*}
   V_0^\ast = \frac{\nu\gamma N}{\nu N+ \beta I^\ast}  \left( 1- \exp\left[-\frac{\beta(1-\omega)P}{N}I^\ast\right]\right)^{-1} I^\ast.
\end{equation*}
The equation for the total population $N=S^\ast+I^\ast+R^\ast+\int_0^P V^\ast(x)\, dx$ renders an equation for $I^\ast$
\begin{align}
    N &= \frac{\gamma N}{\nu N + \beta I^\ast} I^\ast
    + \frac{\gamma}{\alpha} I^\ast
    + I^\ast
    + \frac{\nu\gamma N^2}{\beta(1-\omega)(\nu N+ \beta I^\ast)}. \notag
\intertext{Introducing $z:=I^\ast/N$, $1/R:=\gamma/\beta$, $\lambda:=\frac{\nu}{\beta}$ and $\delta:=\frac{\gamma}{\alpha}$, we get}
    1 &= \frac{z/R}{\lambda+z} + (1+\delta)z + \frac{\lambda}{R(1-\omega)(\lambda+z)}, \notag
\intertext{or}
    0 &= z^2 + \left[\frac{1-R}{R(1+\delta)}+\lambda\right]z + \left[\frac{1}{R(1-\omega)}-1\right]\frac{\lambda}{1+\delta} =:f(z)\;. \label{E:quadEqn_for_z}
\end{align}
Therefore, we arrive at a quadratic equation for $z$, that needs to be solved for $0< z \le 1$ to obtain a meaningful endemic equilibrium.
If $f(0)\cdot f(1) <0$, then there exists exactly one endemic equilibrium $0<z<1$. For $f(1)$ we get
\begin{align*}
    f(1) &= 1+ \frac{1-R}{R(1+\delta)}+\lambda +  \left[\frac{1}{R(1-\omega)}-1\right]\frac{\lambda}{1+\delta}, \\
    & = \frac{1}{1+\delta} \left[ (1+\lambda)(1+\delta) - \frac{R-1}{R} - \lambda(1-\frac{1}{R(1-\omega)}\right], \\
    & = \frac{1}{1+\delta} \left[ \delta(1+\lambda) + \frac{1}{R} + \frac{\lambda}{R(1-\omega)}\right] > 0.
\end{align*}
Thus $f(1)>0$ always and the only option to get exactly one endemic equilibrium is $f(0)<0$ or $R>\frac{1}{1-\omega}\ge 1$.

In order to obtain two endemic equilibria,  the following conditions should be satisfied:
\begin{enumerate}
\item $f(0) >0$, i.e. $R<\dfrac{1}{1-\omega}$ or $\dfrac{1}{R}>1-\omega$.
\item The minimum for $f$ to be located at $0<z_m<1$. Then we have
\begin{equation*}
    z_m := - \frac{1}{2} \left[\frac{1-R}{R(1+\delta)}+\lambda\right],
\end{equation*}
and hence
\begin{gather*}
    1- (2+\lambda)(1+\delta) < \frac{1}{R} < 1 - \lambda(1+\delta).
\end{gather*}
The above two conditions together imply $\omega>\lambda(1+\delta)$ for the possibility to have two endemic equilibria.
\item $f(z_m)<0$. We have
\begin{equation*}
    f(z_m) = \left[\frac{1}{R(1-\omega)}-1\right] \frac{\lambda}{1+\delta} - \frac{1}{4}
    \left[\frac{1-R}{R(1+\delta)}+\lambda\right]^2.
\end{equation*}
Here the condition $f(z_m)<0$ reads as
\begin{gather*}
    4\frac{\lambda(1+\delta)}{1-\omega} \left(\frac{1}{R} -(1-\omega)\right) < \left( \frac{1}{R} - (1-\lambda(1+\delta))\right)^2.
\end{gather*}
\end{enumerate}
So, if $\omega>\lambda(1-\delta)$, there is some region for $R$, where two endemic equilibria could be possible. In terms of $\frac{1}{R}$, the lower bound of this region is given by $\max \left( 1-\omega,\; 1-(2+\lambda)(1+\delta)\right)$ and the upper bound equals to the intercept of the linear function $4\frac{\lambda(1+\delta)}{1-\omega} \left(\frac{1}{R} -(1-\omega)\right)$ and the parabola $\left( \frac{1}{R} - (1-\lambda(1+\delta))\right)^2$.
\\~\\
\begin{center}
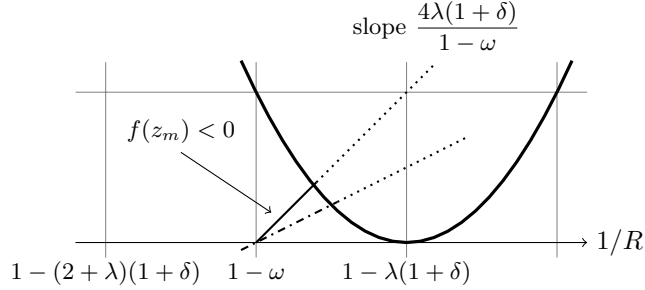

\begin{tikzpicture}[xscale=2, yscale=2]
    \draw[very thin,color=gray] (-2.2,-0.1) grid (1.2,1.2); 
    \draw[->] (-2.2,0) -- (1.2,0) node[right] {$1/R$}; 
    \draw[domain=-1.1:1.1, very thick] plot (\x, {\x*\x});
    \draw[domain=-1:-0.6, thick] plot (\x, {\x+1});
    \draw[domain=-0.6:0.2, thick, dotted] plot (\x, {\x+1}) node[above]{\small slope $\dfrac{4\lambda(1+\delta)}{1-\omega}$};
    \draw[domain=-1.1:-0.46, thick, dashdotted] plot (\x, {(\x+1)/2});
    \draw[domain=-0.46:0.4, thick, dotted] plot (\x, {(\x+1)/2});
    \node at (0,-0.2) {\small $1-\lambda(1+\delta)$};
    \node at (-1,-0.2) {\small $1-\omega$};
    \node at (-2,-0.2) {\small $1-(2+\lambda)(1+\delta)$};
    \draw[->] (-1.5,0.6) node[above]{\small $f(z_m)<0$} -- (-0.9,0.2);
\end{tikzpicture}
\captionof{figure}{Graphical sketch of the condition $f(z_m)<0$ for the existence of two equilibria.}
\end{center}

\section{SIRSV-ODE-model with discrete vaccination age}

Modeling the vaccination age $x$ as a discrete variable, we arrive at a coupled ODE system. Let $k=0,\dots, P$ denote the discrete vaccination ages equal to the time steps (days) elapsed since vaccination and let $V_k$ denote the number of individuals of vaccination age $k$. Re-vaccination is applied to all individuals of $P$ time units, i.e.~individual of vaccination age $P-1$ get re-vaccinated at the next time step and have vaccination age $0$ afterwards. The waning effect of the vaccination is described by a decreasing sequence $\omega_k$. Then the SIRSV-model can now be written as
\begin{subequations}
\label{E:ODEsys-SIRSV}
\begin{align}
    S' &= - \beta S \frac{I}{N} - \nu S  + \alpha R, \label{E:S}\\
    I' &= \beta \frac{I}{N} \left( S + \sum_{k=0}^{P-1} (1-\omega_k) V_k\right) - \gamma I,
    \label{E:I}\\
    R' &= \gamma I - \alpha R, \label{E:R} \\
    V_0' &= \left( 1-\beta (1-\omega_{P-1}) \frac{I}{N} \right) V_{P-1} - V_0 +\nu S, \label{E:V0} \\
    V_k' &= \left( 1-\beta (1-\omega_{k-1}) \frac{I}{N} \right) V_{k-1} - V_k
                \qquad \text{for } k=1,\dots, P-1 \; \label{E:Vk}.
\end{align}
\end{subequations}
For later reference, we introduce the state vector $Z=(S,I,R,V_0,\dots, V_{P-1})\in \R^{P+3}$ of the above ODE system written shortly as $Z'=\Phi(Z)$. In this setting the total population $N=\sum_{j=1}^{P+3} Z_j = S+I+R+\sum_{k=0}^{P-1} V_k$
remains constant.

Again, NPIs can be included in the transmission rate $\beta$, see Eqn.~\eqref{E:NPI}

The disease free equilibrium (DFE) $Z^0$ of the ODE-model equals to $I^0=0$ and hence $R^0=0$. The remaining compartments differ depending on the vaccination rate $\nu$. In case of no vaccinations, i.e.~$\nu=0$, we get no vaccinated individuals at all, i.e.~$V_k^0=0$ for all $k=0,\dots, P$ and all individuals remain susceptible, $S=N$. In the other case $\nu>0$, the susceptible and not vaccinated compartment gets empty, i.e.~$S^0=0$ and due to Eqns.~\eqref{E:V0},\eqref{E:Vk}, the vaccinated distribute evenly over all compartments, i.e.~$V_k^0=\frac{N}{P}$ for all $k=0,\dots, P-1$.

To compute the basic reproduction number of our model system, we use the next-generation matrix approach by van den Driessche and Watmough \cite{van2002reproduction}. The ODE system~\eqref{E:ODEsys-SIRSV} has only one infectious compartment $I$ and hence
\begin{align*}
    \calR_0 &= \frac{\beta}{\gamma N} \left( S^0 + \sum_{k=0}^{P-1} (1-\omega_k) V_k^0\right)\;.
\intertext{In case of $\nu=0$, i.e.~$S^0=N$ and $V_k^0=0$ for all $k$, this simplifies to the well-known basic reproduction number of the standard SIR-model}
    \calR_{0,\nu=0} &= \frac{\beta}{\gamma}\;.
\intertext{In the other case $\nu>0$, we have $S^0=0$ and $V_k^0=\frac{N}{P}$. Hence}
    \calR_{0,\nu>0} &= \frac{\beta}{P \gamma} \sum_{k=0}^{P-1} (1-\omega_k)\;.
\intertext{In case of constant $\omega_k \equiv \omega$, this simplifies to}
    \calR_{0,\nu>0} &= \frac{\beta}{\gamma} (1-\omega).
\end{align*}
If $\omega=1$, i.e. a perfect vaccine, $\calR_0=0$ and for $\omega=0$, i.e.~an effectless vaccine, we get $\calR_0=\frac{\beta}{\gamma}$; the basic reproduction number of the classical SIR--model.

To analyze endemic equilibria $Z^\ast$ with $I^\ast>0$, we again assume $\omega_k=\omega$ for all $k$ for simplicity. We proceed similar to the PDE case; solving the ODE for $R$ yields
\begin{equation*}
    R^\ast = \frac{\gamma}{\alpha} I^\ast.
\end{equation*}
Solving the recursion for $V_k$ yields
\begin{equation*}
    V_k^\ast = c^k V_0^\ast,
\end{equation*}
where $c:= 1 - \frac{\beta (1-\omega)}{N} I^\ast$.
Note, that this recursion is only meaningful, i.e. non-oscillating and bounded, if $0<c<1$. Hence $z =\frac{I^\ast}{N}<\frac{1}{(1-\omega)\beta}$.
\\
Solving the ODE for $V_0'=0$ allows the computation of $S^\ast$:
\begin{equation*}
    S^\ast = \frac{1}{\nu} V_0^\ast (1-c^P).
\end{equation*}
The equation for $I'=0$ yields either the disease free equilibrium $I^0=0$ or the following equation
\begin{gather*}
     \frac{1}{\nu} V_0^\ast (1-c^P) + (1-\omega) V_0^\ast \sum_{k=0}^{P-1} c^k = \frac{\gamma}{\beta} N,
\intertext{i.e.}
    V_0^\ast (1-c^P) \left (\frac{1}{\nu}+ \frac{1-\omega}{1-c}\right) = \frac{\gamma}{\beta} N.
\end{gather*}
And finally, the condition $N=S^\ast+I^\ast +R^\ast+\sum_{k=0}^{P-1} V_k^\ast$ yields
\begin{gather*}
    N = \frac{V_0^\ast}{\nu} (1-c^P) + \frac{\alpha+\gamma}{\alpha} I^\ast + V_0^\ast \sum_{k=0}^{P-1} c^k
    = V_0^\ast (1-c^P) \left( \frac{1}{\nu} + \frac{1}{1-c} \right) + \frac{\alpha+\gamma}{\alpha} I^\ast.
\end{gather*}
Inserting $V_0^\ast (1-c^P) = \frac{\gamma}{\beta} N \left(\frac{1}{\nu} + \frac{1-\omega}{1-c}\right)^{-1}$ yields
\begin{gather*}
    N = \frac{\gamma}{\beta} N \left(\frac{1}{\nu} + \frac{1-\omega}{1-c}\right)^{-1} \left( \frac{1}{\nu} + \frac{1}{1-c} \right) + \frac{\alpha+\gamma}{\alpha} I^\ast.
\end{gather*}
Setting $z=I^\ast/N$ we get
\begin{gather*}
    1 = \frac{\gamma}{\beta} \frac {\nu + \beta (1-\omega)z}{(1-\omega)(\beta z+\nu)} + \frac{\alpha+\gamma}{\alpha} z.
\intertext{or using the same scaled constants $R=\frac{\beta}{\gamma}$, $\lambda=\frac{\nu}{\beta}$ and $\delta=\frac{\gamma}{\alpha}$ as in the PDE case}
    1 = \frac{\lambda+(1-\omega)z}{R(1-\omega)(\lambda+z)}+(1+\delta) z.
\end{gather*}
Again, the same quadratic equation as~\eqref{E:quadEqn_for_z} in the PDE case for $z$ is to be solved.

To analyze the linear stability of the DFE $Z^0=(0,0,0,\frac{N}{P},\dots, \frac{N}{P})$ we introduce the Jacobian $\calJ=(J_{ij})=\frac{\partial \Phi_i}{\partial Z_j}\Big|_{Z^0}$ of the ODE system~\eqref{E:ODEsys-SIRSV} at the DFE $Z^0$. Then
\begin{equation}\label{E:Jacobian}
    \calJ = \begin{pmatrix}
   -\nu & 0 & \alpha & 0 & 0& 0& \cdots & 0 \\
   0 & J_{II} & 0 &0 &0 &0 & \cdots & 0 \\
   0 & \gamma & -\alpha & 0 & 0& 0& \cdots & 0 \\
   \nu & J_{0I} & 0 & -1 & 0&0 & \cdots & 1 \\
   0 & J_{1I} & 0 & 1 & -1 &0 & \cdots& 0 \\
   0 &J_{2I} & 0 & 0 & 1 &-1 &  & 0 \\
   \vdots & & \vdots  &  & \ddots &\ddots& \ddots & \vdots \\
   0 &J_{(P-1)I} & 0 & 0 & \cdots& 0 &1 & -1 \\
\end{pmatrix},
\end{equation}
where $J_{II}= \frac{\beta}{N}\sum_{k=0}^{P-1} (1-\omega_k) V_k^\ast -\gamma = \frac{\beta}{P}\sum_{k=0}^{P-1} (1-\omega_k) -\gamma$ and
$J_{kI} = - \frac{\beta}{N} (1-\omega_k) V_k^\ast = - \frac{\beta}{P} (1-\omega_k)$.

The characteristic polynomial of $\calJ$ equals
\begin{equation*}
    \det(\calJ-\lambda E) = (J_{II}-\lambda) \cdot (-\alpha-\lambda)\cdot (-\nu-\lambda) \cdot \det(\calC_P-\lambda E),
\end{equation*}
where $\calC_P$ denotes the cyclic matrix
\begin{equation*}
    \calC_P = \begin{pmatrix}
        -1 & 0 &0 &\cdots & 1 \\
        1 & -1 & 0 & \cdots & 0 \\
        0 & 1 & -1 & \ddots & 0 \\
        \vdots & \ddots & \ddots & \ddots & \vdots \\
        0 & \cdots & 0 & 1 & -1
    \end{pmatrix} \in \R^{P\times P}.
\end{equation*}
Using Laplace with respect to the first row yields
\begin{equation*}
    \det(\calC_P-\lambda E) = (-1-\lambda)^P+(-1)^{P+1} = (-1)^P \left( (1+\lambda)^P-1\right)\;.
\end{equation*}
Hence
\begin{equation}
    \det(\calJ-\lambda E) = (-1)^P\cdot (J_{II}-\lambda) \cdot (\lambda+\alpha) \cdot (\lambda+\nu) \cdot \left( (1+\lambda)^P-1\right)\;.
\end{equation}
The last factor has the trivial root $\lambda=0$ and all other roots are in the negative half plane $\Real(\lambda)<0$. If $J_{II}<0$, then $\cal J$ has a zero eigenvalue and all other eigenvalues in the negative half plane. The appearance of a zero eigenvalue is obvious due to the conservation of the total population. The condition $J_{II}<0$ is equivalent to $\calR_{0,\nu>0}<1$. Therefore the DFE $Z^0$ is linearly stable, if the basic reproduction number $\calR_{0,\nu>0}<1$.

\begin{table}[htbp]
    \centering
    \caption{Description of the parameters of the SIRSV-system \eqref{E:ODEsys-SIRSV}.}
\begin{tabular}{lrp{.5\textwidth}} \label{Tab:Parameters}
\renewcommand{\arraystretch}{1.5}
Parameter & Value & Remark \\ \hline
$N$ & 1000  & Total population\\
$S(0)$ & 995 & Initial susceptibles\\
$I(0)$ & 5 & Initial infected\\
$R(0)$ & 0 & Initial recovered\\
$V_k(0)$ & 0 & No vaccinated at the beginning\\
$\gamma$ & 0.1 & Recovery period of $10$ days\\
$\beta$ & 0.23 & Infection rate\\
$\alpha$ & 0.005 & Loss of immunity after $200$ days\\
$\nu$ & 0.01 & Vaccination rate\\
$P$ & 90 & Vaccination update after 3 months\\
$\omega_{k}$ & $e^{-k/W}$ & Waning efficiency of the vaccine, $W=60$
\end{tabular}
\end{table}

In the sequel we will work with the ODE formulation of the SIRSV-model.

\section{Numerical study of the epidemiological SIRSV-model}

This section is devoted to a comprehensive   numerical investigation
of the dynamical behavior of the SIRSV-ODE-model \eqref{E:ODEsys-SIRSV}
when selected control parameters are varied. To this end, we
will use path-following (continuation) methods implemented by the continuation software COCO \cite{dankowicz2013}. Based on the
MATLAB environment, this platform is focused on the numerical solution of continuation problems covering a wide range of analysis
and detection capabilities available in classical continuations
packages, such as AUTO \cite{auto97} and MATCONT \cite{matcont}. In
this work two such COCO functionalities will be employed, namely,
continuation and bifurcation detection of equilibria of parameter-dependent smooth odes, and that of multi-segment periodic
solutions of piecewise-smooth systems, as will be explained in
detail later.

\begin{figure}
\centering
\psfrag{S}{\large$S(t)$}\psfrag{I}{\large$I(t)$}\psfrag{Vks}{\large$V_{k}(t)$}\psfrag{t}{\large$t$
\scriptsize[days]}
\includegraphics[width=0.5\textwidth]{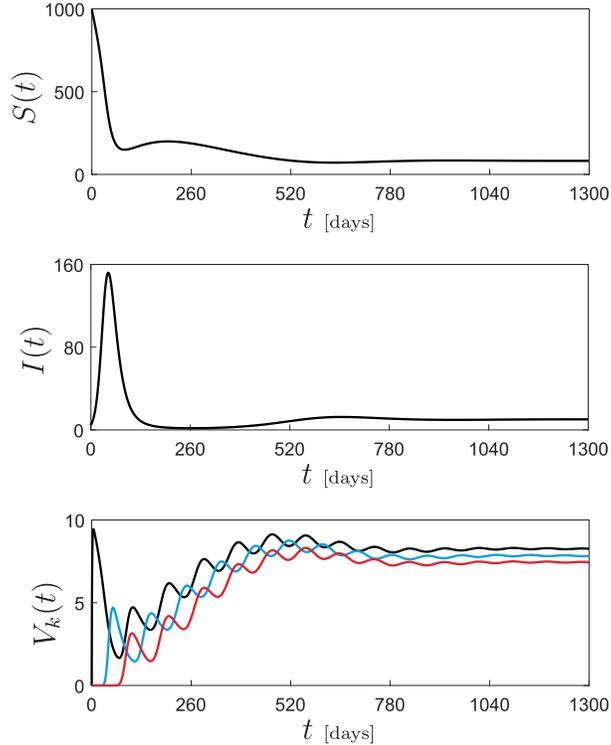}
\caption{Simulation of the epidemiological SIRSV-model
\eqref{E:ODEsys-SIRSV}, calculated using the parameter values and
initial conditions provided in Table \ref{Tab:Parameters}, with
$\omega_{k}=0.5$ constant (no waning effect). The last graph shows time
series of $V_{k}$-compartments for $k=0$ (black), $k=45$ (blue) and
$k=89$ (red).}\label{fig-sol-ini}
\end{figure}

\subsection{Investigation of parameter-dependent equilibria of the
model}

For the numerical study of the epidemiological SIRSV-model
\eqref{E:ODEsys-SIRSV}, it is important to note first that some parameters and variables presented in Table \ref{Tab:Parameters} span different orders of magnitude. This variation often impacts the precision and reliability of the results because of the set tolerances for determining convergence and estimating errors in COCO. This imbalance can make the computations less responsive to certain variations while being excessively sensitive to others. Hence, we re--scale the SIRSV model \eqref{E:ODEsys-SIRSV} as shown below
$$S\leftarrow\frac{S}{N},\mbox{ }\mbox{ }\mbox{ }I
\leftarrow\frac{I}{N},\mbox{ }\mbox{ }\mbox{
}V_{k}\leftarrow\frac{V_{k}}{N},\mbox{ }\mbox{ }\mbox{
}k=0,1,\ldots,P-1.$$ Moreover, since the overall population $N$ remains constant, we will reduce the dimension of the system
via the equation $R=N-S-I-\sum\limits^{P-1}_{k=0}V_{k}$.
Furthermore, although the numerical computations will be implemented
using the re-scaling introduced above, the outcome will be presented
in the original scale for better interpretation of the results.

\begin{figure}
\centering
\psfrag{a}{\large(a)}\psfrag{b}{\large(b)}\psfrag{c}{\large(c)}
\psfrag{d}{\large(d)}\psfrag{beta}{\LARGE$\beta$}\psfrag{nu}{\LARGE$\nu$}
\psfrag{R0}{\Large$\calR_{0}$}\psfrag{eq}{\Large$I$}
\psfrag{S}{\large$S(t)$}\psfrag{I}{\large$I(t)$}\psfrag{t}{\large$t$
\scriptsize[days]}
\includegraphics[width=\textwidth]{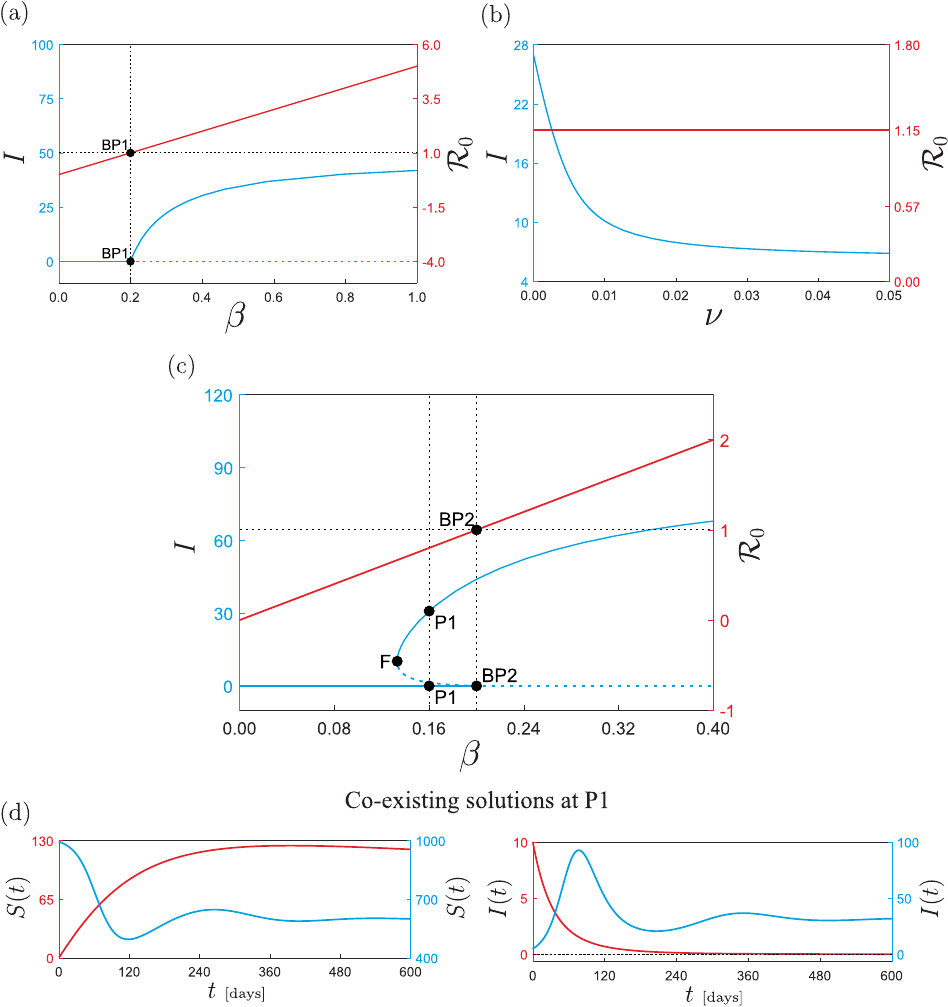}
\caption{Numerical continuation of equilibrium points of system
\eqref{E:ODEsys-SIRSV} with respect to $\beta$ and $\nu$. Panels (a)
and (b) are calculated for the parameter set given in Fig.\
\ref{fig-sol-ini}, while panel (c) is obtained for the same
parameters, except for $\nu=0.0003$ and $\alpha=0.01$. Panels (a)--(c) display the infection compartment behavior at equilibrium on the left vertical axis (blue), the second vertical axes (right, in red) represent the basic reproduction number $\calR_{0}$. In these plots, solid blue lines indicate stable equilibrium branches, and dashed blue lines represent unstable equilibrium branches. Throughout the calculations, various critical points are identified, including the branching points BP1 and BP2 at $\beta=0.2$, and the fold bifurcation F at $\beta\approx0.13303$. Panel (d) presents time series for the system \eqref{E:ODEsys-SIRSV}  at the test point P1 ($\beta=0.16$), where two stable solutions co--exist.}\label{fig-1par-cont-eq}
\end{figure}

To start our study, we will set the parameter values of the
epidemiological SIRSV-model \eqref{E:ODEsys-SIRSV} according to
Table \ref{Tab:Parameters}, but considering $\omega_{k}=0.5$
constant, that is, we will assume a perfect vaccine with a fixed
efficiency (no waning effect). Under this consideration, the
dynamical response of system \eqref{E:ODEsys-SIRSV} obtained via
direct numerical integration is presented in Fig.~\ref{fig-sol-ini}. This diagram displays the transient response for the following compartments of the model: $S$, $I$ and $V_{k}$
($k=0,45,89$). As can be observed in the diagram, with the chosen parameter values, the system exhibits a damped oscillatory behavior that eventually stabilizes over time into an endemic equilibrium, that is, a steady state where the disease persists within the system. A notable
feature of this dynamical response is the pulse-like outbreak of the
disease observed during the first $100$ days, reaching a peak of
around $152$ infections. After this, the infections consistently
decrease owing to the effectiveness of the vaccination campaigns,
although at the end such pharmaceutical intervention is not able to
eradicate completely the disease. Furthermore, the
$V_{k}$-compartments show a biologically consistent behavior, in
that they activate themselves in the expected sequence, according to
the proposed re-vaccination scheme.

Next, we will use numerical continuation methods to explore how variations in certain parameters affect the equilibrium and validate some of the analytical results derived in earlier sections. Specifically, our analysis will discuss the response of the endemic equilibrium to continuous variations of the infection rate $\beta$ and vaccination rate $\nu$. The results from the path-following analysis are presented in Fig.~\ref{fig-1par-cont-eq}, panels (a)-(c). In these figures, the left vertical axis (in blue) shows the simulated infected compartment with respect to variations of the model paramters $\beta$ or $\nu$. The second vertical axis (right hand side, in red) displays the according reproduction n umber $\calR_0$.

Fig.\ref{fig-1par-cont-eq}(a) shows the equilibria as the infection rate $\beta$ changes. The diagram reveals that the reproduction number $\calR_0$ is below $1$, resulting in a stable disease-free equilibrium, as long as the infection rate is below the brancing pojnt BP! at $\beta_0.2$. This is represented by the solid horizontal branch in the figure. As $\beta$ increases, $\calR_{0}$ also rises, crossing 1 from below at thee branching point BP1.

At this point, the an endemic equilibrium emegeres and the disease-free equilibrium gets unstable through a forward bifurcation. For larger values of the infection rate, the disease persists and the number of infections grows as the infection rate increases. A different scenario arises when the vaccination rate $\nu$ is adjusted, as shown in Fig.\ref{fig-1par-cont-eq}(b). As can be observed in the
picture, for $\nu$ between 0 and 0.01 a strong reduction of the
infections is achieved, thus showing the effectiveness of the
proposed vaccination campaign. However, a more intensive (and
expensive) vaccination policy produces no significant additional
reduction, as can be verified during the numerical continuation of
equilibria. From a mathematical perspective, this can be explained
observing the response of the basic reproduction number
$\calR_{0}$, which does not depend on $\nu$ (see red curve in Fig.~\ref{fig-1par-cont-eq}(b)). This is due to the fact that for the
disease-free equilibrium the susceptible compartment gets empty, and
hence there is no new individuals to vaccinate at rate $\nu$, as
explained in the previous section. This means that vaccination alone
cannot eliminate the disease, or even reduce the level of infected
humans below an arbitrarily chosen level, if required.

In the previous section, our analytical study of the proposed
epidemiological SIRSV-model revealed the possibility of having a
single endemic equilibrium (as the scenario discussed above) or two
coexisting endemic equilibria. In order to verify this prediction,
we will choose suitable parameter values according to the analytical
calculations presented earlier in order to confirm numerically the
expected behavior. The outcome is presented in Fig.~\ref{fig-1par-cont-eq}(c), which was obtained for the same set of
parameters as before, except for $\nu=0.0003$ and $\alpha=0.01$,
chosen according to our analytical calculations. The continuation
process detects of a backward bifurcation at BP2
($\beta=0.2$), which produces precisely the expected scenario,
namely, the presence of coexisting endemic equilibria, one stable
and one unstable. At high infection rates, the system exhibits a branch of endemic equilibria, shown in blue in the figure. This branch becomes unstable at a fold bifurcation F, occurring at approximately $\beta=0.13303$. This fold bifurcation occurs, when a second endemic equilibrium appears, i.e.~when the quadratic equation~\eqref{E:quadEqn_for_z} allows for two real solutions corresponding to these two possible equilibria. Setting the discriminant of the quadratic equation~\eqref{E:quadEqn_for_z} equal to zero yields
$$\beta_{\mbox{\tiny fold}}=\gamma-\nu(1+\delta)\pm 2\sqrt{\gamma\nu(1+\delta)}.$$
For the parameters chosen in Fig.~\ref{fig-1par-cont-eq}(c), the larger of the two solution is given by $\beta\approx 0.133$, the fold bifurcation $F$. At this point, the endemic equilibrium branch shifts in the direction of increasing parameters and ends at the backward bifurcation BP2. Here, the basic
reproduction number $\calR_0$ crosses transversally the critical
value $1$ from below when the parameter increases over BP2.

It is important to note in the previous discussion that having $\calR_0<1$ does not necessarily guarantee the disease to be eradicated. Between F and BP2, we find a bistable region, and the initial condition determines, whether, the system either approachs the disease-free or the endemic equilibrium. This is illustrated in Fig.~\ref{fig-1par-cont-eq}(d) using the parameter value $\beta=0.16$, which lies within the bistability range. From a practical standpoint, this presents a critical scenario where, even with $\calR_0<1$, the sudden introduction of sufficient virus carriers into the population could drive the system toward an endemic equilibrium with high infection levels. Furthermore, once the system enters the endemic branch, reducing the infection rate $\beta$ to bring $\calR_0$ below 1 (as seen at BP2) is not enough to eradicate the disease. The infection rate must be reduced below the fold bifurcation F, at which point the disease-free equilibrium becomes globally stable.

\subsection{Disease control considering contact restrictions}

\begin{figure}
\centering \psfrag{a}{\large(a)}\psfrag{b}{\large(b)}
\psfrag{I}{\large$I(t)$}\psfrag{t}{\large$t$
\scriptsize[days]}\psfrag{rho}{\large$\rho(t)$}\psfrag{Vks}{\large$V_{k}(t)$}
\psfrag{Imax}{$I_{\mbox{\tiny max}}$}
\includegraphics[width=\textwidth]{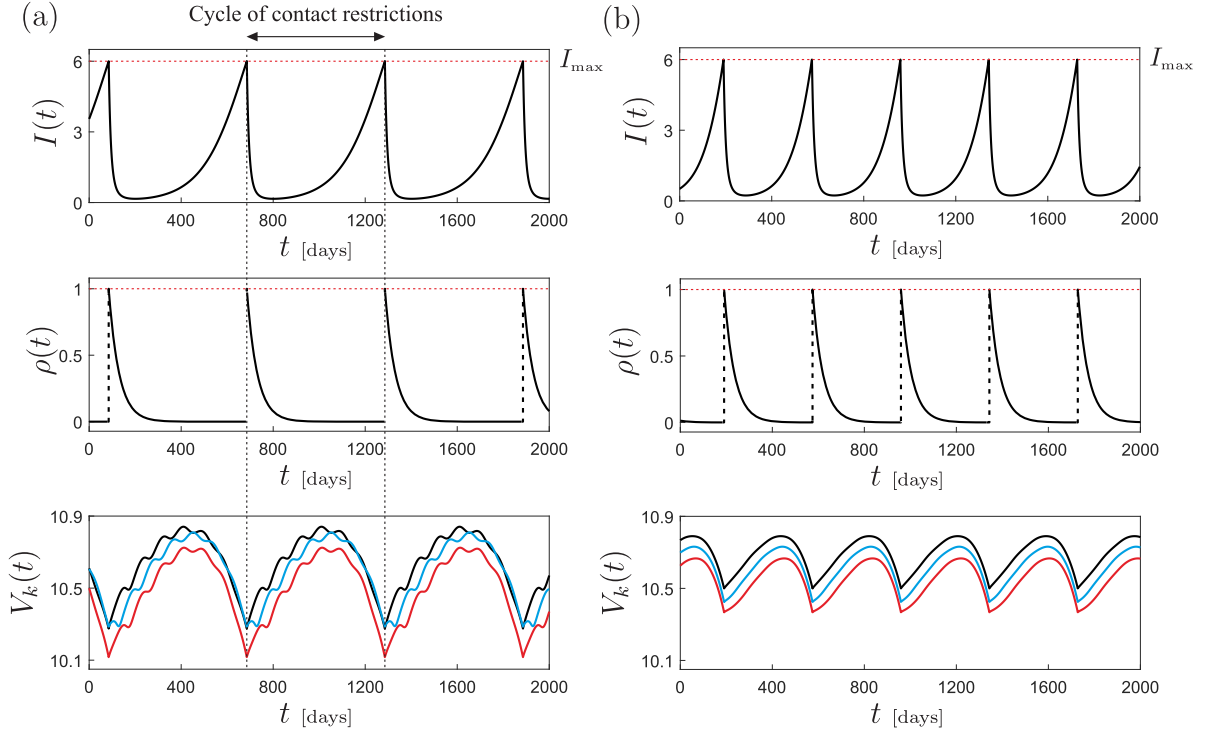}
\caption{Simulation of the epidemiological SIRSV-model
\eqref{E:ODEsys-SIRSV} subject to the contact restrictions
\eqref{E:NPI} and \eqref{eq-rho}, calculated for the parameter
set provided in Table \ref{Tab:Parameters} with $\eta=\frac{1}{45}$
and $I_{\mbox{\tiny max}}=6$. The $V_{k}$-compartments follow the
same color code as in Fig.~\ref{fig-sol-ini}. Panel (a) is
calculated considering the waning effect, while panel (b) is
computed for $\omega_{k}=0.5$ constant (no
waning).}\label{fig-sol-control}
\end{figure}

As the SARS-CoV-2 (COVID-19) pandemic in 2019 showed, one effective
way to contain the disease is to impose contact restrictions
(lockdowns) on the population, in combination to massive vaccination
campaigns. In the previous section, our numerical study
established that for the proposed epidemiological SIRSV-model
\eqref{E:ODEsys-SIRSV} vaccination alone is not enough to achieve an
effective control of the number of infections, and therefore
additional control strategies should be employed. Similar to the
COVID-19 case, we will introduce contact restrictions in the model
according to the functional response \eqref{E:NPI}. Here, the
function $\rho$ is described by
\begin{equation}\label{eq-rho}
\begin{cases}
\rho(t)=1, & I(t)=I_{\mbox{\tiny max}},\mbox{ }\mbox{ }\mbox{
}I'(t)>0,\\
\rho'(t)=-\eta\rho(t), & \mbox{otherwise},
\end{cases}
\end{equation}
with $0\leq\rho(0)\leq1$ and $t\geq0$. From \eqref{E:NPI} one can see that the
main role of the function $\rho$ is to control the transmission rate
$\beta$, where a full lockdown corresponds to $\rho=1$ and no
contact restrictions to $\rho=0$. In model \eqref{eq-rho} it is
assumed that full lockdown is applied when the number of infections
reaches a predefined critical level $I_{\mbox{\tiny max}}$. After
this, the contact restrictions are relaxed gradually as the time
progresses, assuming a simple exponential decay (which can of course
follow any other decaying pattern, if required). The pace at which
the restrictions are relaxed is controlled by the parameter
$\eta>0$. A small value (say close to 0) indicates that the
restriction policies are strict, while $\eta$ high means that the
lockdowns do not last for too long. A numerical simulation of this
control scheme is presented in Fig.~\ref{fig-sol-control} showing
two cases: with waning vaccine efficiency ($\omega_{k}$ variable,
panel (a)) and fixed vaccine efficiency ($\omega_{k}$ constant,
panel (b)).

\subsection{Monitor and cost functions}

Notice that according to the control scheme proposed in the previous
section the peak of infections are restricted via the parameter
$I_{\mbox{\tiny max}}$. Every time the infection compartment reaches
this value from below full lockdowns are activated, due to which the
number of infections decreases, as can be observed in the simulation
shown in Fig.~\ref{fig-sol-control}. Therefore, one monitor
function that we will consider in our investigation is the following
\begin{equation}\label{eq-mess}
I_{\mbox{\tiny AVG}}=\frac{1}{T_{0}}\int\limits_{0}^{T_{0}}I(t)\,dt,
\end{equation}
where a bounded periodic response with period $T_{0}$ is assumed.
This gives the average number of infections per period, which is a
quantity that is independent from the peak infections
$I_{\mbox{\tiny max}}$. That is, for a fixed $I_{\mbox{\tiny max}}$
one may observe different levels of $I_{\mbox{\tiny AVG}}$ when
other parameters are adjusted, as will be seen later.

Another factor that will be examined in our study is the cost associated with the proposed control scheme. Specifically, we will consider
two costs as follows
\begin{equation}\label{eq-costfun}
P_{\mbox{\tiny
COST}}=\frac{1}{T_{0}}\int\limits_{0}^{T_{0}}\frac{\rho(t)}{\frac{I(t)}{N}}\,dt\mbox{
}\mbox{ }\mbox{ }\text{(political cost)},\mbox{ }\mbox{ }\mbox{
}V_{\mbox{\tiny
COST}}=\frac{\nu}{T_{0}}\int\limits_{0}^{T_{0}}S(t)\,dt\mbox{
}\mbox{ }\mbox{ }\text{(vaccination cost)}.
\end{equation}
The first term can be interpreted as a measure of the political cost
produced by the lockdown policies. When $\rho(t)$ decreases slowly
and remains close to $1$ (see previous section) it means that the
contact restrictions are strict, and therefore the political cost
should increase. If, however, such policy is applied during a
disease outbreak (i.e.~high $I(t)$) then the lockdowns can be well
justified among the population and hence the political cost reduces.
On the contrary, if strict lockdowns are applied ($\rho(t)$ close to
$1$) during low level of infections ($I(t)$ close to $0$), then the
political cost grows, because such policy would seem completely
unnecessary. On the other hand, the second term in
\eqref{eq-costfun} represents the cost of vaccination campaigns. It
is taken to be the product of the average number of susceptible
individuals during one period and the vaccination rate, which is an
indicator of how many individuals are vaccinated in average and
gives us insight as to how the associated costs behave.

\subsection{Numerical study of the proposed disease control strategy}

This section will present a numerical investigation of the
dynamical behavior of the epidemiological SIRSV-model
\eqref{E:ODEsys-SIRSV} considering contact restriction policies
given by \eqref{E:NPI} and \eqref{eq-rho}, with $\beta_{0}=0.23$
fixed. To this end, we will use the COCO-capabilities for
numerical continuation of multi-segment periodic solutions for
piece-smooth systems, since model \eqref{E:ODEsys-SIRSV} in
combination with \eqref{E:NPI} and \eqref{eq-rho} falls within this
category. For our investigation, we will consider the two cases
displayed in Fig.\ \ref{fig-sol-control}: panel (a) including waning
efficiency of the vaccine (decreasing $\omega_{k}$) and panel (b)
corresponding to $\omega_{k}=0.5$ fixed (no waning).

\begin{figure}
\centering
\psfrag{a}{\large(a)}\psfrag{b}{\large(b)}\psfrag{c}{\large(c)}\psfrag{d}{\large(d)}
\psfrag{e}{\large(e)}\psfrag{f}{\large(f)}
\psfrag{nu}{\LARGE$\nu$}\psfrag{eta}{\LARGE$\eta$}
\psfrag{Iavg}{\large$I_{\mbox{\tiny
AVG}}$}\psfrag{Pcost}{\large$P_{\mbox{\tiny
COST}}$}\psfrag{Vcost}{\large$V_{\mbox{\tiny COST}}$}
\includegraphics[width=\textwidth]{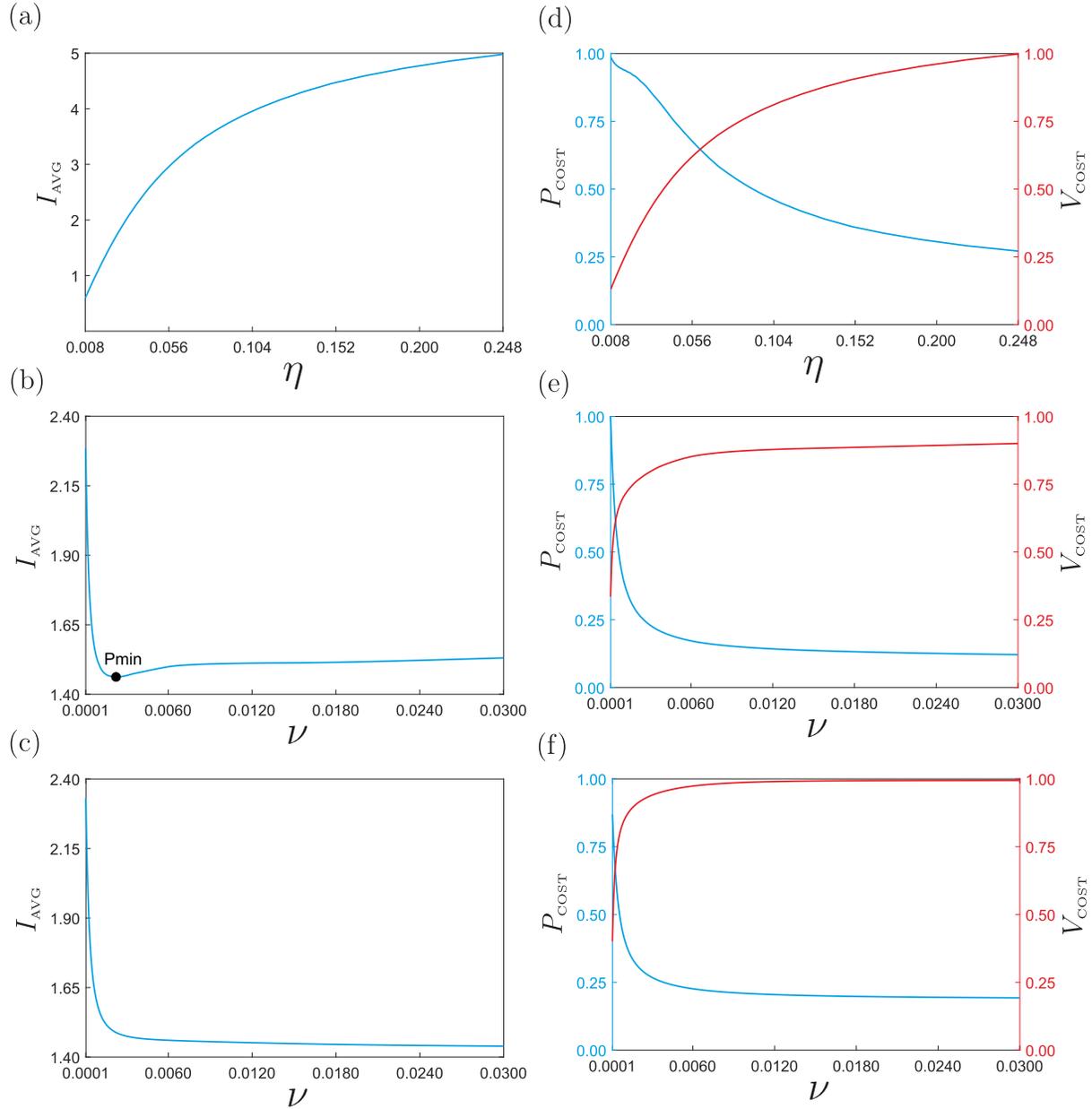}
\caption{Numerical continuation of the periodic solution shown
in Fig.~\ref{fig-sol-control} with respect to $\eta$ and $\nu$.
Panels (a)--(c) show the behavior of the average infection occurrences per period. Panels (d)--(f) display the variation of the
political cost (left vertical axis, blue) and vaccination cost
(right vertical axis, red) as the control parameter changes. All
calculations are performed using the periodic response in Fig.~\ref{fig-sol-control}(a) (waning effect), except for the curves in
panels (c) and (f), which correspond to the solution in Fig.~\ref{fig-sol-control}(b) (no waning). In these diagrams, both
political and vaccination costs are normalized between $0$ and $1$ for
better comparison.}\label{fig-1par-cont-po}
\end{figure}

Let us consider first the case of waning vaccine efficiency. The
numerical continuation of the resulting periodic response with
respect to the control parameter $\eta$ is presented in Fig.~\ref{fig-1par-cont-po}(a). As explained earlier, this parameter
determines how fast the contact restrictions are relaxed. Small
values of this parameter indicate that the lockdowns are strict and
only lifted slowly in time. This effect is reflected on the behavior
of the average number of infections $I_{\mbox{\tiny AVG}}$, in an
expected and consistent manner. In addition, Fig.~\ref{fig-1par-cont-po}(d) displays the behavior of the control costs
defined in \eqref{eq-costfun}. As can be seen in this figure, the
political cost and vaccination cost present decreasing and
increasing behavior, respectively. The political cost reduces as the
parameter $\eta$ increases, since larger $\eta$ means soft contact
restrictions, which do not pose any discomfort to the population.
The vaccination cost, however, increases due to the increasing
number of average infections that after some time goes back to the
susceptible population via the recovery compartment $R$ (see
\eqref{E:ODEsys-SIRSV}), and therefore the number of vaccinated
individuals grows.

A different scenario is found when the vaccination rate $\nu$ is
considered as control parameter. Fig.~\ref{fig-1par-cont-po}(b)
shows the response of the average number of infections
$I_{\mbox{\tiny AVG}}$ as the parameter varies. Similar to the
equilibrium case (see Fig.~\ref{fig-1par-cont-eq}(b)) a strong
reduction of average infections can be achieved over a short
parameter window. Beyond that, the value of $I_{\mbox{\tiny AVG}}$
stabilizes regardless of how strong the vaccination rate becomes. As
for the control costs, one can see in Fig.~\ref{fig-1par-cont-po}(e) that again vaccination and political costs
present different behaviors. As expected, the vaccination cost
increases when $\nu$ grows. The political cost, on the contrary,
increases significantly when the vaccination becomes smaller. This
is due to the fact that as the vaccination campaigns weaken, the
only remaining action to keep $I(t)$ below $I_{\mbox{\tiny max}}$ is
the contact restrictions through the function $\rho$, and, as
discussed earlier, stronger restrictions increase the incurred
political cost. This means that in order to keep the political costs
under reasonable levels, both vaccination and lockdowns should be
combined in a suitable manner.

Another feature observed during the numerical continuation of the periodic solutions with waning vaccine efficiency is the presence of
an optimal level of vaccination rate. This is found for
$\nu\approx0.00227$ (point labeled Pmin), see Fig.~\ref{fig-1par-cont-po}(b). Here, a local minimum of average number
of infections $I_{\mbox{\tiny AVG}}$ is found, and one question that
can be tackled here is whether the presence of this minimum is
affected by the type of vaccine, namely, a perfect vaccine with
constant efficiency (see solution in Fig.~\ref{fig-sol-control}(b))
and a vaccine that loses effectiveness over time (waning effect).
Therefore, we will also perform the numerical continuation of the
system response shown in Fig.~\ref{fig-sol-control}(b) (perfect vaccine)
with respect to the vaccination rate $\nu$, see Fig.~\ref{fig-1par-cont-po}(c). Here, it can be observed that there is no
minimum anymore, which indicates that the type of vaccine plays an
important role in choosing suitable vaccination rates, in terms of
minimizing average number of infections, and not incurring in
unnecessary increasing vaccination costs, see Figs.~\ref{fig-1par-cont-po}(b), (c), (e) and (f).

\section{Conclusion}
\label{Sec:Conc}

The SIRSV-model, described in this paper, provides a comprehensive framework for understanding the dynamics of infectious diseases in the presence of waning vaccine efficiency and periodic re-vaccination. By incorporating both vaccination and re-vaccination processes, along with NPIs, the model is able to capture the critical elements required for effective disease control in real-world scenarios. Our analysis demonstrates that while vaccination plays a vital role in reducing the spread of infections, the waning efficiency of vaccines presents a significant challenge to long-term immunity. The periodic re-vaccination strategy, as modeled here, helps to mitigate this effect, but does not eliminate the risk of endemic disease. Through equilibrium and stability analysis, we identified the conditions under which the disease-free equilibrium is stable and the circumstances that lead to the persistence of endemic equilibria.

One of the principal findings from our numerical simulations is the importance of combining vaccination with timely NPIs, such as contact restrictions. Our results show that NPIs can significantly reduce the average number of infections, especially when vaccine efficiency wanes over time. However, the model also demonstrates a trade-off between the costs of NPIs, vaccination, and the effectiveness of disease control. The optimization of these control measures is of critical importance for the design of policies that minimize both the number of infections and the economic and political costs associated with interventions.

Future work can focus on refining the model by incorporating more realistic vaccination schedules, heterogeneous populations, and age-structured dynamics. Furthermore, the cost functions introduced in this paper can be further developed to include more detailed economic and social factors, thus providing a more comprehensive tool for public health decision-making \cite{Moore2021}, \cite{wijaya2021}.

\bibliographystyle{ieeetr}
\bibliography{sirbg.bst}

\end{document}